\documentclass[%
 reprint,
 amsmath,amssymb,
 aps,
prb,
]{revtex4-2}

\usepackage{graphicx}
\usepackage{dcolumn}
\usepackage{bm}
\usepackage{xcolor}

\newcommand{\updownarrows}{\mathbin\uparrow\hspace{-.0em}\downarrow}

\begin{document}

\title{Symmetry-Breaking Magneto-Optical Effects in Altermagnets}
\author{Jiuyu Sun}
\author{Yongping Du}
\email{njustdyp@njust.edu.cn}
\author{Erjun Kan}
\email{ekan@njust.edu.cn}
\affiliation{Department of Applied Physics, Nanjing University of Science and Technology, Nanjing 210094, China}
\affiliation{MIIT Key Laboratory of Semiconductor Microstructure and Quantum Sensing, Nanjing University of Science and Technology, Nanjing 210094, China}

\date{\today}

\begin{abstract}
The recently discovered altermagnets (AMs), hosting momentum-dependent spin splitting and vanishing net magnetization, have attracted intensive attention for their promising application in novel spintronics.
However, limited by facility and material constraints, experimentally distinguishing them from conventional antiferromagnets (AFMs) remains a challenge, which hinders the high-throughput screening for AM candidates.
Here, we predict strain-mediated magneto-optical responses in AMs, which can  serve as a universal and experimentally accessible strategy for efficient identification of AMs.
Symmetry analysis reveals that uniaxial strain can selectively breaks rotation or mirror symmetries in AMs while preserving $PT$ symmetry in AFMs, thereby activating distinct linear magneto-optical responses (e.g., optical absorption and Kerr rotation) unique to AMs.
First-principles calculations across prototypical systems---including semiconducting V$_2$Se$_2$O monolayer and metallic CrSb bulk---show that the strain-induced optical signatures are significant enough for conventional optical measurements. Our work establishes a rapid, non-invasive characterization methodology for altermagnetism across material platforms, accelerating its exploration for spin-based technologies.
\end{abstract}

\maketitle



{\it Introduction.}---Altermagnetism, a recently discovered magnetic order featuring collinear spin arrangements with vanishing net magnetization, has rapidly emerged as a cornerstone for next-generation spintronics due to its unique symmetry-protected spin-splitting phenomena\cite{Smejkal2020,Ma2021,Smejkal2022a,Zhu2024,Krempasky2024,Song2025AltermagReview,Fender2025}. Unlike conventional antiferromagnets (AFMs), altermagnets (AMs) host momentum-dependent spin-polarized electronic states, enabling unprecedented opportunities for spin transport and manipulation\cite{Zhang2024,Bai2024,Gu2024,Wang2025,Li2025,Chen2025,Tan2025}. While theoretical advances have greatly expanded the family of candidate AMs\cite{Yu2020,Liu2022,Sodequist2024,Zeng2024a,Qi2024,Leeb2024,Pan2024,Che2025,Che2025a}, direct experimental verification remains limited to a handful of materials, including MnTe\cite{Mazin2023,Krempasky2024,Hariki2024a,Lee2024}, CrSb\cite{Ding2024,Zeng2024,Reimers2024,Zhou2025}, RuF$_4$\cite{Milivojevic2024}, Rb$_{1-\delta}$V$_2$Te$_2$O\cite{Zhang2025}, KV$_2$Se$_2$O\cite{Jiang2025}, and the contentious candidate RuO$_2$\cite{Kessler2024,Liu2024}.
One of the bottlenecks is experimentally identifying altermagnetism in an arbitrary material, particularly for efficiently differentiating AMs from conventional AFMs given their shared vanishing macroscopic magnetization. 
Current identification approaches in experiment face critical barriers: 
Direct observation with spin-resolved angle-resolved photoemission spectroscopy\cite{Zeng2024,Ding2024,Lee2024,Fedchenko2024} and indirect probe with X-ray magnetic circular dichroism (XMCD)\cite{Hariki2024a,Hariki2024,Fedchenko2024}, muon spin relaxation\cite{Kessler2024}, and high-end (time-resolved) optical spectroscopy\cite{Zhu2023,Gray2024,Ma2025-SHG} often demand stringent sample conditions or large-scale/intricate experimental facilities;  Electrical transport measurements based on anomalous Hall effect (AHE)\cite{GonzalezBetancourt2023} are generally hindered by contact resistance and Fermi-level misalignment, while detectable signals emerge only in AMs with specific symmetry conditions tied to N{\'{e}}el vector orientation\cite{Smejkal2020,Bose2022,Feng2022,Zhou2025,Regmi2025,Sheoran2025}.
These challenges highlight the urgent need for a methodology that eliminates facility/sample constraints, bypasses material dependence and enables rapid material screening---a capability vital for accelerating the discovery and validation of AM candidates.

Here in this work, we propose a general and accessible methodology to rapidly distinguish AMs from conventional AFMs by harnessing strain-engineered magneto-optical responses.
By symmetry analysis and a crystal-field model, we demonstrate that controlled uniaxial strain selectively breaks the symmetry-protected compensated spin-texture in AMs, while retaining the $PT$ symmetry in AFMs, therefore inducing pronounced differences in their signatures of optical response. First-principles calculations reveal that magneto-optical effects (optical absorption and/or MOKE), even induced by a weak strain, are significant enough for experimental detection in prototypical AMs ranging from 2D limit (monolayer V$_2$Se$_2$O) to bulk (CrSb).
Our work establishes an alternative strategy for identifying altermagnetism across diverse material systems, significantly accelerating the discovery and functional exploration of this appealing magnetic phase.

{\it Symmetry considerations and microscopic interpretation.}---We start our symmetry-based analysis with optical conductivity within the framework of independent-particle approximation with spin-orbit coupling (SOC) included:
\begin{equation}\label{conductivity}
\sigma_{\alpha\beta}(\omega)  \propto \sum_{\mathbf{k}} \sum_{n,m} \frac{f_{m\mathbf{k}} - f_{n\mathbf{k}}}{E_{m\mathbf{k}} - E_{n\mathbf{k}}} \frac{\langle\psi_{n\mathbf{k}}|\hat{\upsilon}_{\alpha}|\psi_{m\mathbf{k}}\rangle \langle\psi_{m\mathbf{k}}|\hat{\upsilon}_{\beta}|\psi_{n\mathbf{k}}\rangle}{E_{m\mathbf{k}} - E_{n\mathbf{k}} - (\hbar\omega + i\eta)}
\end{equation}
where the subscripts $\alpha,\beta \in \{x, y, z\}$ denote Cartesian coordinates, the $\psi_{n\mathbf{k}}$ and $E_{n\mathbf{k}}$ are respectively the wavefunction and energy at band index $n$ and momentum $\mathbf{k}$ with a distribution function of $f_{n\mathbf{k}}$, $\hat{v}_{\alpha,\beta}$ are velocity operators, $\hbar\omega$ is the photon energy, and $\eta$ is an adjustable energy smearing parameter (0.1 eV for all the calculations in this work). 
The $\sigma_{\alpha\beta}(\omega)$ relates to dielectric tensor via $\varepsilon_{\alpha\beta}(\omega)=\delta_{\alpha\beta}+i\frac{4\pi}
{\omega}\sigma_{\alpha\beta}(\omega)$. The diagonal terms correspond to the optical properties like linear absorption (Im[$\varepsilon_{\alpha\alpha}$]), which can be approximately decomposed into the contributions from spin-up and spin-down channels in the collinear spin systems, namely, $\varepsilon^{\uparrow}_{\alpha\alpha}$ and $\varepsilon^{\downarrow}_{\alpha\alpha}$, respectively. Notably $\varepsilon^{\uparrow}_{\alpha\alpha} = \varepsilon^{\downarrow}_{\alpha\alpha}$ holds rigorously in AFMs, due to their completely compensated spin texture at every $\mathbf{k}$, i.e., $E^{\uparrow}_{n\mathbf{k}} = E^{\downarrow}_{n\mathbf{k}}$. Remarkably, this degeneracy ($\varepsilon^{\uparrow}_{\alpha\alpha} = \varepsilon^{\downarrow}_{\alpha\alpha}$) persists even in AMs despite spin-splitting, as guaranteed by the summation over counterpart $\mathbf{k}$ and $\mathbf{k'}$-points ($E^{\uparrow}_{n\mathbf{k}} = E^{\downarrow}_{n\mathbf{k'}}$) in Eq.~\ref{conductivity}.
Thus, it is hard to distinguish AMs from AFMs via their spin-degenerate optical absorption spectra even with circularly polarized light.

On the other hand, the off-diagonal components of $\sigma_{\alpha\beta}(\omega)$ and $\varepsilon_{\alpha\beta}(\omega)$, enabled by SOC, govern magneto-optical phenomena such as the Faraday and Kerr effects. For example, the polar MOKE could be computed as\cite{Wu2019,Molina-Sanchez2020,Wu2022}
\begin{equation}
    \theta_K + i\eta_K = \frac{-\varepsilon_{\alpha\beta}(\omega)}{(\varepsilon_{\alpha\alpha}(\omega)-1)\sqrt{\varepsilon_{\alpha\alpha}(\omega)}}
\end{equation}
where $\theta_K$ and $\eta_K$ denote the rotation angle and ellipticity, respectively, with magnetization perpendicular to the reflection surface and parallel to the plane of incidence ($\alpha\beta$-plane).
Crucially, the $PT$ symmetry requirement $PT\sigma_{\alpha\beta} = -\sigma_{\alpha\beta}=0$ strictly prohibits MOKE in AFMs. 
In contrast, AMs break $T$ symmetry, allowing finite $\sigma_{\alpha\beta}$ which explains the experimentally observed magneto-optical signals\cite{Hariki2024a}.
However, broken $T$ symmetry alone does not universally guarantee non-vanishing  $\sigma_{\alpha\beta}$ in all AMs\cite{Zhou2021,Kimel2024,Hariki2024,Sheoran2025}. Specific symmetry combinations---such as the coexistence of spin symmetry $C_2$ and spatial mirror symmetry $M$ in CrSb\cite{Reimers2024,Zhou2025} and many two-dimensional materials\cite{Sheoran2025}---can enforce $\sigma_{\alpha\beta}=0$.
This fundamental constraint, originating from the Néel vector ($\bm L$) orientation-dependent symmetry conditions, indicates that intrinsic optical properties obtained through conventional probes are insufficient to unambiguously distinguish AMs from AFMs.

\begin{figure}[h]
\includegraphics[width=8 cm]{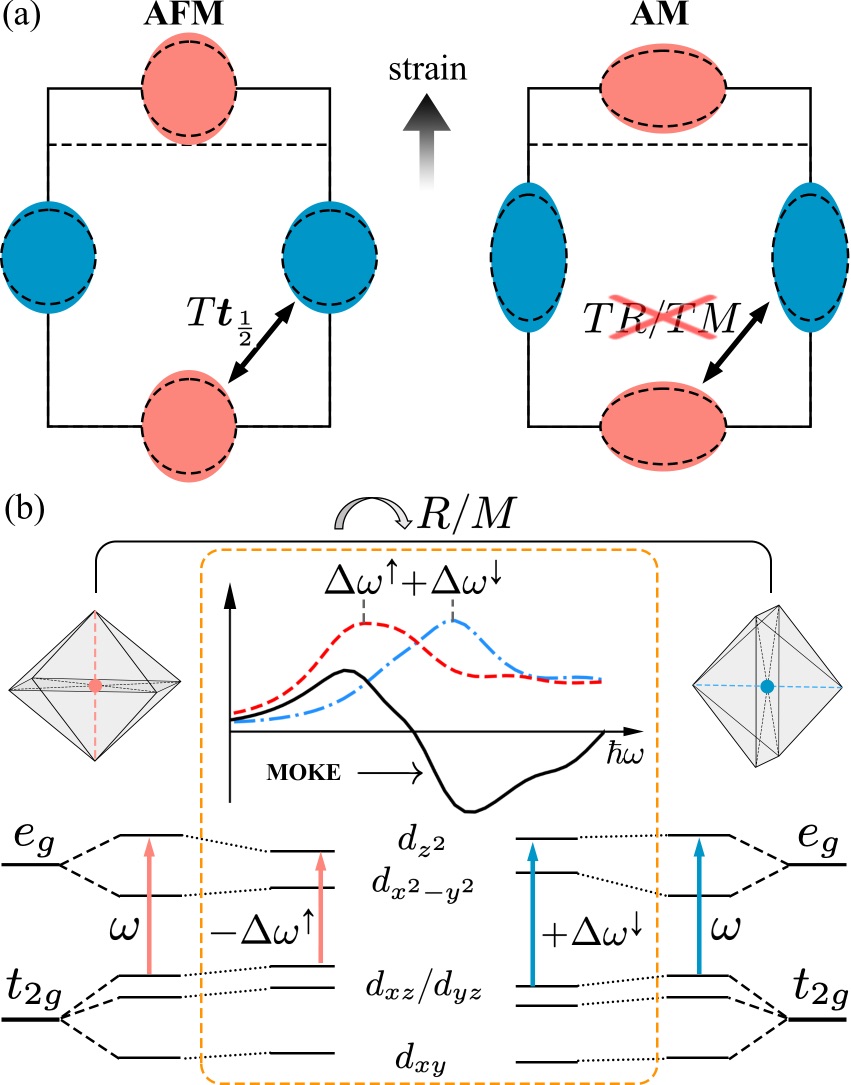}
\caption{\label{fig1} (a) Schematic diagram of the effects by a uniaxial strain in AFMs and AMs. The magnetization isosurfaces of different spins indicated by ovals (spheres) with different colors. The dashed lines indicate the unstrained lattices and magnetization isosurfaces. (b) Demonstration of effects from strain on the crystal field levels and optical responses in AMs with $R/M$ spatial symmetry. The central atom of spin-up (spin-down) and direction of its $d_{z^2}$ orbitals are indicated by the red (blue) dots and dashed lines. The parts inside of the orange dashed-line box illustrates the changes made by the strain. The $d-d$ optical transitions are shown by the red (blue) arrows. The schematic of the optical absorption in spin-up (spin-down) channel Im[$\varepsilon^{\uparrow}_{\alpha\alpha}$] (Im[$\varepsilon^{\downarrow}_{\alpha\alpha}$]) is shown by red dashed (blue dot-dashed) line, along with solid black line for the Kerr rotation $\theta_K$.}
\end{figure}

To this end, it requires introducing additional perturbations that selectively break rotational or mirror ($TR/TM$) symmetries in AMs while keeping the $PT$ symmetry in AFMs. Here we employ the experimentally deployable uniaxial strain field. As exemplified by a 2D AFM with lattice vectors $\bm a$ and $\bm b$ (Figure~\ref{fig1}a),  the spin-opposite sublattices are connected by $T$ combined with half-lattice translation $\bm{t}_{\frac{1}{2}}=(\frac{\bm a}{2},\frac{\bm b}{2})$. When applied by an uniaxial strain along an arbitrary direction, the $PT$ symmetry is preserved only with $\bm{t}_{\frac{1}{2}}$ turning into another $\bm{t}'_{\frac{1}{2}}$, keeping the spin degeneracy in AFM.
On the other hand, the spin-sublattices (ovals in Figure~\ref{fig1}a) in AMs are connected by the rotational/mirror ($R/M$) operations, which are broken by applying strain along non-symmetry-adapted directions (e.g., vertical direction in Fig.~\ref{fig1}a). This distortion leads the ovals of opposite spins into different shapes with distinct orbital hybridization geometries, manifesting as fundamentally distinct optical responses compared to strain-free AMs or strained AFMs.

\begin{figure*}[htp]
\centering
\includegraphics[width=17 cm]{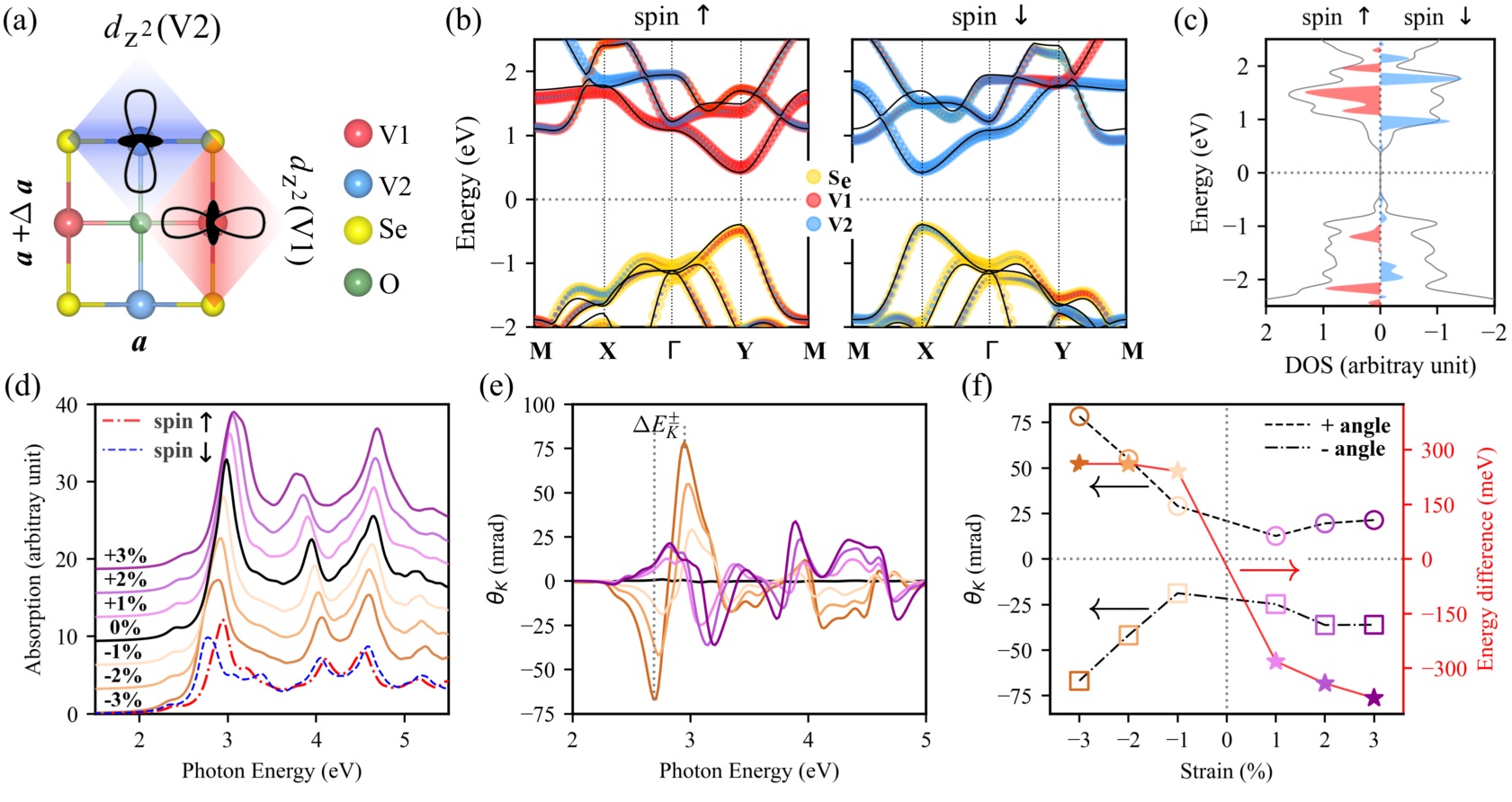}
\caption{(a) The top view of crystal structure of strained monolayer V$_2$Se$_2$O with V1 and V2 atoms carrying spin-up and spin-down magnetic moments, respectively. (b) The spin-resolved band structures of unstrained (black solid lines) and -3\% strained (colored) monolayers, where the -3\% strained bands are projected onto the $d$-orbitals of V1 (V2) and $p$-orbital of Se with the weights proportional to the radii of the colored circles. Energy zero is at the middle of direct band gap. (c) The spin-resolved DOS (solid gray lines) of 3\% strained V$_2$Se$_2$O, where the difference of DOS between spin-up and spin-down channels 
($\Delta \mathrm{DOS}^{\updownarrows}$) 
is plotted in red (blue) color if spin-up (-down) DOS is the majority at certain energy levels. (d) The strain-dependent diagonal optical spectra Im($\varepsilon$) by BSE for the light polarization along [110]. (e) Strain-dependent Kerr rotation angle $\theta _{K}$. (f) The tendencies of strain-dependent largest positive and negative $\theta_{K}$ around 3 eV, and the difference between their photon energies, $\Delta E^{\pm} _{K}$, as illustrated in (e). }\label{fig2}
\end{figure*}

The effects of the strain field on the optical response in AMs can be understood via the crystal-field model as demonstrated in Fig.~\ref{fig1}b. 
We consider a prototypical $d$-wave AM with spin-polarized transition metal centers forming octahedral pairs connected by $R/M$ symmetries. 
While the orientations of two octahedra (orientations of $d_{z^2}$ orbitals) are non-parallel, the $d$ orbitals are energy degenerate in the absence of strain (out of the orange box). When subjected to a non-symmetry-adapted strain, the spin-up (red) and spin-down (blue) octahedra experience opposite lattice distortions, and the energy levels of $d$ orbitals exhibit strain-dependent splittings. For example, applying a strain along the orientation of spin-up octahedron stretches the $d_{z^2}$ orbital in spin-up octahedron, whereas the $d_{z^2}$ orbital in spin-down octahedron is fractionally stretched or even compressed due to the Poisson’s ratio effects.
As a result, energy level of $d_{z^2}$ orbital in spin-up octahedron is shifted downwards more than the one in spin-down octahedron. Similar changes apply to all the components $d$-orbitals of the central atoms, as well as the orbitals of corresponding ligands.
Since the central atoms carry magnetization opposite to each other, the spin-degeneracy of energy levels in the AM is therefore broken, i.e. $E^{\uparrow}_{n\mathbf{k}} \ne E^{\downarrow}_{n\mathbf{k}}$.
For the intra-atomic $d-d$ transitions, e.g. from half-filled $d_{xz}/d_{yz}$ to empty $d_{z^2}$, the transition energy (frequency $\omega$) within its majority spin-up (spin-down) channel decreases (increases) by $\Delta\omega^{\uparrow}$ ($\Delta\omega^{\downarrow}$).
This indicates the strain-induced changes to the optical transitions, $E^{\uparrow}_{m\mathbf{k}}-E^{\uparrow}_{n\mathbf{k}} \ne E^{\downarrow}_{m\mathbf{k}}-E^{\downarrow}_{n\mathbf{k}}$ in Eq.~\ref{conductivity}, and thus the non-degenerate optical absorption spectra ($\varepsilon^{\uparrow}_{\alpha\alpha} \ne \varepsilon^{\downarrow}_{\alpha\alpha}$).
It is also possible to observe a splitting of $\Delta\omega^\uparrow+\Delta\omega^\downarrow$ for the peaks between Im[$\varepsilon^{\uparrow}_{\alpha\alpha}$]  and Im[$\varepsilon^{\downarrow}_{\alpha\alpha}$], as demonstrated in Figure~\ref{fig1}b.
Moreover, thanks to the spin-uncompensated electronic distribution at particular energy levels analogous to the ferromagnets, the spin-majority optical transitions also give rise to a non-zero MOKE with the peak position relevant to the the shifts ($\Delta\omega^\uparrow$ and $\Delta\omega^\downarrow$) of the peaks in Im[$\varepsilon^{\uparrow}_{\alpha\alpha}$]  and Im[$\varepsilon^{\downarrow}_{\alpha\alpha}$] spectra (Figure~\ref{fig1}b). 
Remarkably, this mechanism transcends specific symmetry realizations: any AM will exhibit analogous strain-tunable optical dichroism---a universal fingerprint distinguishing AMs from strain-insensitive AFMs.

{\it Computational verification.}---
Using first-principles calculations, we verify the theoretical analysis above and explore the magnitude of strain-induced magneto-optical effects in diverse prototypes of AMs. 
Density functional theory plus Hubbard $U$ (DFT+$U$) method in Quantum-Espresso code\cite{QE,QE-2017} is adopted for ground state calculations with PBE functional\cite{PBE} and optimized norm-conserving Vanderbilt pseudopotentials\cite{ONCV-psp,PseudoDojo}. The optical properties are computed with Yambo code\cite{marini2009yambo,Yambo2019,Molina-Sanchez2020,yambo-spinor,yambo-CoulombCut,BSE,haydock}, with many-body excitonic effects included (excluded) for the semiconducting (metallic) system.
More computational details can be found in the Supplemental Materials (SM)\cite{supp}.

We start with monolayer V$_2$Se$_2$O (Figure~\ref{fig2}(a)), a prototypical AM not only in the 2D limit but also having a finite band gap according to our previous work\cite{VXO-exciton} and others'\cite{Ma2021,Lin2018,Qi2024}. The unstrained V$_2$Se$_2$O monolayer belongs to the spin group $|C_2||C_{4z}|$, with the collinear magnetic moments on the two V atoms aligning oppositely along the out-of-plane $z$-direction. The V1 (spin-up) atom and V2 (spin-down) atom occupy the central sites of two octahedra surrounded by O and Se atoms, which are linked by a $C_{4z}$ operation and a diagonal mirror operation.
As the symmetry of $|C_2||C_{4z}|$ in this monolayer system leads to vanishing off-diagonal optical conductivity $\sigma_{xy}$ in principle, it is challenging to distinguish its altermagnetism from the almost vanishing Kerr rotation angle in the $xy$-plane (see Figure~\ref{fig2}e).

To break the $C_{4z}$ and diagonal mirror symmetries, we apply a uniaxial strain field along $y$-direction by artificially changing the lattice constant $a$ along $y$-direction into $a+\Delta a$, and leaving the lattice constant along $x$-direction fixed. The magnitude of the strain is defined as $\epsilon = - \Delta a/a$, where a positive (negative) $\epsilon$ refers to a compressive (tensile) strain.
According to the crystal-field interpretation above, the orthogonally oriented octahedra containing V1 (red) and V2 (blue) atoms will experience different distortions, leading to opposite shifts in the energy levels of their respective spin states.
In the band structures for the V$_2$Se$_2$O monolayer under a tensile strain of -3\% (Figure~\ref{fig2}b), we observe roughly opposite shifts of bands in different spins. For example, the first and second lowest conduction bands (CBs) of spin-up around M point, move upward and downward, respectively, while the corresponding spin-down bands exhibit the opposite behavior.
In addition to the CBs (mostly contributed by the $d$-orbitals of V atoms), the valence bands (VBs) contributed by the $p$-orbitals from Se atoms also have spin-dependent response to the strain field, e.g. the top two VBs along $\Gamma$-X and $\Gamma$-Y of spin-up and spin-down, respectively.
To explore the spin-resolved response of the electronic structure in the whole Brillouin zone, we plotted the difference of density of state (DOS) between two spin-channels. i.e., $\Delta \mathrm{DOS}^{\updownarrows} = \mathrm{DOS}^\uparrow - \mathrm{DOS}^\downarrow$. Although the total electron count per spin channel is conserved in this semiconductor, the nonzero $\Delta \mathrm{DOS}^{\updownarrows}$ in Figure~\ref{fig2}c clearly indicates the spin-uncompensated distribution of DOS at specific energy levels. 
Furthermore, one could infer from the $\Delta \mathrm{DOS}^{\updownarrows}$ that, for example, the optical transitions taking place in the bands between -1 eV (VBs) to 1 eV (CBs) in spin-down channel will be prior to the those in spin-up channel.

For a better estimation of optical properties of a 2D finite-gap system, we included the excitonic effects by solving Bethe-Salpeter equation based on a rigid scissors shift of 2.0 eV to the PEB+U wavefunction. The calculated linear optical absorption spectra (Im[$\varepsilon_{\alpha\alpha}$]) of monolayers under various strain fields are plotted in Figure~\ref{fig2}d, which shows a remarkable dependence on the strain field. For instance, the stronger the tensile (compressive) strain, the larger red-shift (blue-shift) will be observed for the highest peak (around 3 eV), along with a larger broadening of the peaks. By calculating the collinear projected optical spectra (Im[$\varepsilon^{\uparrow}_{\alpha\alpha}$] and Im[$\varepsilon^{\downarrow}_{\alpha\alpha}$], dashed lines in Figure~\ref{fig2}d), we confirm that this broadening is attributed to the broken degeneracy of optical absorptions of different spins.
Note that the shift of optical peaks could be induced by the strain field in materials with any magnetic orders, while the broadening or splitting of the peaks is mostly attributed to the broken degeneracy of energy levels of different spins, which could further be detected with light fields of different circular polarization.

In Figure~\ref{fig2}e, we show the MOKE spectra of V$_2$Se$_2$O monolayer under varying uniaxial strain field.
While the unstrained monolayer exhibits negligible Kerr rotation angle ($\theta_{K}$), applying just 1\% tensile strain amplifies $\theta_{K}$ by several orders of magnitude, with the positive and negative angles stranding for right- and left-hand Kerr rotations caused by the energy-resolved uncompensated spin-up and spin-down DOS, respectively. Interestingly, we observe clear inverse double-peak structures in the MOKE spectra around 3 eV with tensile strains. This is consistent with the non-degenerate absorptions around 3 eV between two spin channels. 
Furthermore, we define $\Delta E^\pm_K$ as the difference between the photon energies corresponding to the largest positive and negative $\theta_{K}$ around 3 eV. As shown in Figure~\ref{fig2}f,  the compressive strain field tends to produce larger $\Delta E^\pm_K$, while the tensile strain can generally induce larger $\theta_{K}$. We note that the $\theta_{K}$, even with 1\% tensile strain, is comparable to that in monolayer CrI$_3$\cite{Molina-Sanchez2020}, which allows for detectable signals in experiments.

\begin{figure}[htb]
\centering
\includegraphics[width=8.5 cm]{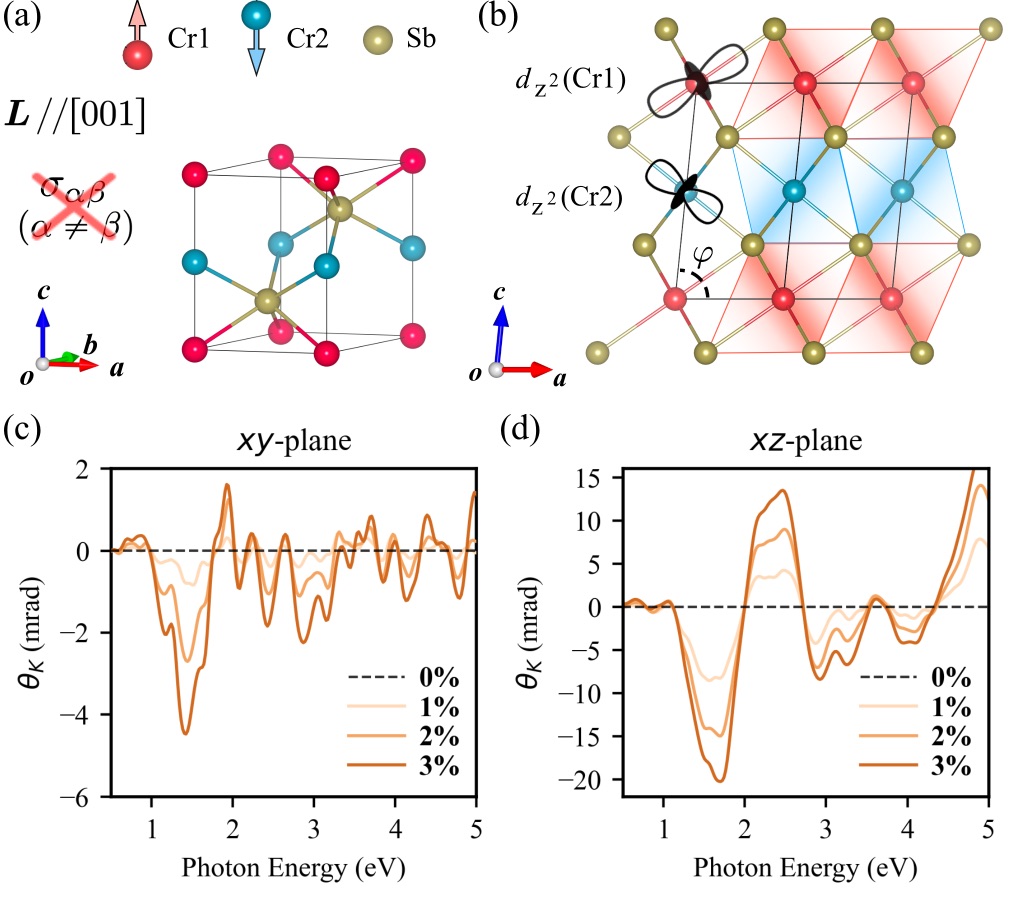}
\caption{The crystal structures of (a) unstrained and (b)  strained (side-view) CrSb, where the red (blue) arrows on Cr1 (Cr2) atoms indicate the direction of their magnetic moments. Within (c) $xy$-plane and (d) $xz$-plane, strain-dependent Kerr rotation angle $\theta_K$.}\label{fig3}
\end{figure}

Exemplified by the hexagonal CrSb, we demonstrate that our strategy is also applicable to the systems with other symmetries. As reported in previous works\cite{Reimers2024,Zhou2025}, CrSb has a space group of $P6_3/mmc$ and a N{\'{e}}el vector along [001] direction, with the spin-up and spin-down octahedra connected by a $C_{6z}$ rotation and a $M_z$ mirror (see Figure~\ref{fig3}(a))\cite{Ding2024}. Owing to the anti-symmetric nature of $\sigma_{xy}$, MOKE rotation is strictly prohibited under multiple generalized mirror symmetries in CrSb bulk, including the magnetic mirror symmetry, glide mirror symmetry and glide magnetic mirror symmetry\cite{Zhou2025}.

To overcome this limitation, a uniaxial strain applied along a direction neither aligned with the $C_{6z}$ axis nor lying in the $xy$-plane should be effective. Here, we choose a strain along [111] direction. It is simulated by tilting the lattice vector $\bm{c}$ slightly toward [110], and the angle ($\varphi$) between $\bm{c}$ and $\bm{a}$($\bm{b}$) is set from 90$^{\circ}$ to 87$^{\circ}$ with a step of 1$^{\circ}$. As demonstrated in Figure~\ref{fig3}(a), this uniaxial strain unequally tilts the spin-up (red) and spin-down (blue) octahedra, in which the $d_{z^2}$ orbitals of Cr1 and Cr2 atoms are stretched and compressed, respectively. As expected, for the strained CrSb, we found distinguished shifts between the bands of opposite spins (see SM\cite{supp}). The calculated MOKE spectra within $xy$-plane and $xz$-plane ( Figure~\ref{fig3}c and Figure~\ref{fig3}d, respectively) show that the induced rotation signal almost linearly increases with strain strength. 

Notably, another prototypical AM, MnTe bulk\cite{Kriegner2017-MnTe,Krempasky2024,Lee2024,Hariki2024a}, has the same space group of $P6_3/mmc$ but a N{\'{e}}el vector along [110] direction. This enables non-vanishing MOKE under the $M_z$ operation, i.e. $M_z(\sigma_{zy},\sigma_{xz},\sigma_{yx})=(-\sigma_{zy},-\sigma_{xz},\sigma_{yx})$. Thus, an intrinsic MOKE rotation within $xy$-plane exists in MnTe, while other two vanishing components $\sigma_{xz}$ and $\sigma_{yz}$ prevent MOKE signals within $yz$- and $xz$-planes (see SM\cite{supp}). However, since the crystal orientation of arbitrary samples grown on substrates is difficult to control experimentally, polar MOKE measurements may still yield zero signal if the probe light polarization lies in $xz$- or $yz$-plane. By applying the same strain along [110], 
consequently, we observe not only the enhanced MOKE within $xy$-plane, but also the appearance of MOKE signal within $xz$-plane with even larger $\theta_{K}$ values compared to the $xy$-plane (see SM\cite{supp}).


{\it Summary}---
In this work, we propose a universal and experimentally accessible strategy to rapidly distinguish AMs from AFMs by leveraging strain-engineered magneto-optical responses.
The uniaxial strain selectively breaks rotation or mirror symmetries in AMs while preserving $PT$ symmetry in AFMs, thereby activating distinct linear magneto-optical responses (e.g., optical absorption and Kerr rotation) unique to AMs.
First-principles calculations across prototypical systems---including
semiconducting V$_2$Se$_2$O monolayer, metallic CrSb bulk---show the strain-induced optical signatures are remarkable enough for conventional optical measurements.
Note that in all the strained structures calculated above, the magnetic moments are collinearly arranged without canting. This means 
negligible Dzyaloshinskii–Moriya interaction is induced by the weak strain field, which largely preserves the spin configuration in AMs.
Moreover, since the Hall conductivity could also be computed with off-diagonal $\sigma_{\alpha\beta}$ in Eq.~\ref{conductivity} if the photon energy is set to zero, our strategy of selective symmetry-breaking with  uniaxial strain can be generalized to AHE, suggesting its broad utility for accelerating the investigation of altermagnetism for spin-based technologies.

\begin{acknowledgments}
This work was financially supported by the Ministry of Science and Technology of the People’s Republic of China (No.2022YFA1402901), the National Natural Science Foundation of China (NSFC No.T2125004, No.U24A2010,  No.12274228 and No.22303041), the NSF of Jiangsu Province (No.BK20230908), Fundamental Research Funds for the Central Universities (No.30922010102, No.30922010805 and No.30923010203), and funding (No.TSXK2022D002) as well as a startup grant from the NJUST. The authors acknowledge support from the Tianjing Supercomputer Centre, Shanghai Supercomputer Center, and High Performance Computing Platform of Nanjing University of Aeronautics and Astronautics.
\end{acknowledgments}


\bibliography{strain-AM}

\end{document}